\documentclass{aastex631}
\usepackage{xspace}

\newcommand{\Edotin}{\ensuremath{\dot{E}_\mathrm{in}}}
\newcommand{\Edotout}{\ensuremath{\dot{E}_\mathrm{out}}}
\newcommand{\Sdotin}{\ensuremath{\dot{S}_\mathrm{in}}}
\newcommand{\Sdotout}{\ensuremath{\dot{S}_\mathrm{out}}}
\newcommand{\teff}{\ensuremath{T_\mathrm{eff}}}
\newcommand{\Winternal}{\ensuremath{\dot{W}_\mathrm{internal}}}
\newcommand{\Winternalone}{\ensuremath{\dot{W}_\mathrm{internal,1}}}
\newcommand{\Winternaltwo}{\ensuremath{\dot{W}_\mathrm{internal,2}}}
\newcommand{\about}{\ensuremath{{\sim}}\xspace}

\begin{document}

\journalinfo{Accepted to AAS Journals}

\title{Application of the Thermodynamics of Radiation to Dyson Spheres as Work Extractors and Computational Engines, and Their Observational Consequences}

\newcommand{\PSUAA}{Department of Astronomy \& Astrophysics, 525 Davey Laboratory, The Pennsylvania State University, University Park, PA, 16802, USA}
\newcommand{\PSUCEHW}{Center for Exoplanets and Habitable Worlds, 525 Davey Laboratory, The Pennsylvania State University, University Park, PA, 16802, USA}
\newcommand{\PSETI}{Penn State Extraterrestrial Intelligence Center, 525 Davey Laboratory, The Pennsylvania State University, University Park, PA, 16802, USA}

\author[0000-0001-6160-5888]{Jason T.\ Wright}
\affil{\PSUAA}
\affil{\PSUCEHW}
\affil{\PSETI}

\begin{abstract}

I apply the thermodynamics of radiation to Dyson spheres as machines that do work or computation, and examine their observational consequences.  I identify four properties of Dyson spheres that complicate typical analyses: globally, they may do no work in the usual sense; they use radiation as the source and sink of energy; they accept radiation from a limited range of solid angle; and they conserve energy flux globally. I consider three kinds of activities: computation at the Landauer limit; dissipative activities, in which the energy of a sphere's activities cascades into waste heat, as for a biosphere; and ``traditional'' work that leaves the sphere, such as radio emission. I apply the Landsberg formalism to derive efficiency limits in all 3 cases, and show that optical circulators provide an ``existence proof'' that greatly simplifies the problem and allows the Landsberg limit to be plausibly approached. I find that for computation and traditional work, there is little to no advantage to nesting shells (as in a ``Matrioshka Brain''); that the optimal use of mass is generally to make very small and hot Dyson spheres; that for ``complete'' Dyson spheres we expect optical depths of several; and that in all cases the Landsberg limit corresponds to a form of the Carnot limit.  I explore how these conclusions might change in the face of complications such as the sphere having practical efficiencies below the Landsberg limit (using the endoreversible limit as an example); no use of optical circulators; and swarms of materials instead of shells.  

\end{abstract}

\section{Introduction}

\subsection{The Kinds of SETI and Prior Work on Dyson Spheres}

There are five primary pillars of modern SETI: radio SETI \citep{cocconi_searching_1959,OZMA}, optical SETI \citep{Schwartz61}, solar system SETI \citep{Bracewell60}, waste heat SETI \citep{dyson60}, and exoplanetary SETI \citep{Campbell06}.  Of these, radio SETI is by far the most developed, followed by optical SETI.  The challenge of detecting the interactions of light with terrestrial exoplanetary surfaces and atmospheres makes the last of these an understandably immature field, but waste heat and solar system SETI have a heritage just as old as radio and optical SETI, and remain relatively undeveloped.

The premise of waste heat SETI is that life and technology, almost by definition, exploit free energy gradients.  In particular, life can exploit these gradients and local resources to reproduce, and ``intelligent'' life can overcome many energy and resource limitations through the application of technology \citep{GHAT1}. This means that, in principle, technological life might be able to expand and grow to exploit an almost arbitrarily large amount of energy, up to the absolute limit: the total luminosity if a nearby star. Because energy must be conserved, the star's luminosity must be expelled from the system as waste heat, and so one might be able to measure the total energy use of such technology by looking for such heat.

\citeauthor{dyson60} was partly motivated by the development of infrared detector technology, and imagining what might be found once astronomers could search the infrared sky. He correctly surmised that there would be many confounding sources, especially stars with circumstellar dust, and that the search for ``infrared stars'' would have broad astrophysical implications, regardless of whether any Dyson spheres were found. He also correctly pointed out that finding excess infrared emission from a star would hardly be dispositive evidence of alien life, and that some other traces of technology would need to be found to close the case.

\citet{GHAT1} and \citet{Wright2020_Dyson} provide a thorough discussion of the history of waste heat SETI, including the limited number of searches to date.  The primary challenge is that waste heat will presumably be found at mid-infrared wavelengths, which are best studied above Earth's atmosphere.  A search for such emission thus requires an all-sky midinfrared space mission, of which there have only been three: \textit{IRAS}, \textit{WISE}, and \textit{AKARI}.  \citet{Carrigan09a} provided the first thorough search using \textit{IRAS}, \citet{GHAT3} provided the first limits from \textit{WISE} for extragalactic sources, and \citet{Suazo2022} provided the first limit for nearby stars.

Searches are somewhat hampered, however, by the lack of an underlying theory of the waste heat of circumstellar technological material that can predict its observational consequences.  \citet{GHAT2} presented the AGENT formalism for parameterizing the fraction of starlight reprocessed as waste heat and its approximate effects on stellar and galactic spectral energy distributions.  \citet{Wright2020_Dyson} developed a model for spherically symmetric distributions of circumstellar material,\footnote{That work contains at least two errors: Eq.~25 should read $s = (R_*/R)^2$ and the solid angle subtended by the star is then $2\pi(1-\sqrt{1-s^2})$; and it is missing the factor of $\frac43$ in the rate of computation described later in this work.}  connecting the AGENT parameters to the physical properties of the material, and \citet{Huston2022} determined the effects of such material on the surface and evolutionary properties of the star, making the model fully self-consistent.

These models remain simple and general, however: they generally assume spherical symmetry, assume that material has a characteristic orbital distance from the star, ignore radiative interactions among the material, and do not provide strong guidance regarding the typical temperatures, distributions, or optical depths of the material. To guess at such properties, one must typically invoke specific motivations or purposes for the material.

\subsection{Proposed Motivations and Purposes of Dyson Spheres}

\citet{dyson60}'s original conception of an ``artificial biosphere'' around a star, later dubbed a ``Dyson sphere'' by \citet{kardashev64}, was loosely sketched as a logical limit of a species' expansion into space.  Later, Dyson acknowledged that capturing all of a star's luminosity to maintain a vast habitat was just one possible motivation for such a project; in \citet{Dyson66} he generalized the argument for why such structures might exist or what their purpose might be:

\begin{quote}
    [T]hink of the biggest possible artificial activities, within limits set only by the laws of physics and engineering\ldots I do not need to discuss questions of motivation, who would want to do these things or why. Why does the human species explode hydrogen bombs or send rockets to the moon? It is difficult to say exactly why. My rule is, there is nothing so big nor so crazy that one out of a million technological societies may not feel itself driven to do, provided it is physically possible.
\end{quote}

With this philosophy, we minimize our consideration of Dyson spheres' reasons to exist beyond the very general idea that they consume large amounts of energy to perform some kind of task. They might arise organically as a species expands into space, and not as part of a grand engineering project, slowly blocking more and more of the star's light the way a growing forest eventually develops a canopy that does the same.

Dyson spheres require an enormous extrapolation from current human technology. Humanity has put very roughly  0.1 km$^2$ of solar panels into Earth orbit, enough to block $\about10^{-19}$ of the light that would have escaped into space.\footnote{From Jonathan McDowell (private communication) based on his comprehensive catalog of human objects in space. With solar panels having $\about20\%$ efficiency, humanity's space technology is thus $K\about0.14$ on Kardashev's scale, where a complete Dyson sphere is $K=2$ \citep{Gray2020b}.} To build a shell around the Sun 1 cm think at 1 au would require roughly an Earth-mass of material.  Being so far beyond our current engineering capabilities, one must wonder what we could possibly say about them, including the the possibility that they could even exist. After all, the technology that could rearrange such huge amounts of mass would presumably employ methods we could not imagine.  

\citet{Dyson66} addressed this by considering only the laws of physics we are reasonably sure are foundational, ignoring engineering practicalities except to establish physical possibility:

\begin{quote}
    I am presenting an existence proof for certain technological possibilities. I describe crude and clumsy methods which would be adequate for doing various things. If there are other more elegant methods for doing the same things, my conclusions will still be generally valid.
\end{quote}

His ``crude and clumsy'' methods included ways to disassemble planets to acquire the huge amounts of mass necessary to block a significant fraction of a star's light.

There are many disagreements in the literature about the form a Dyson sphere would take.  \citet{badescu95} analyzed the thermodynamics of Dyson spheres under certain assumptions about how they harvest energy, and attempted to calculate a minimum size for them given that they are biospheres. \citet{badescu2000} discuss and analyze many uses of a Dyson sphere, defining 3 classes of ``stellar engines,'' including a class A engine that propels the star, and a class B engine that performs work. The latter is the primary focus of this paper (Dyson spheres may serve other purposes as well). Others have argued they would be extremely cold to maximize the work they can do (for instance \citet{Lacki16}).  The proper thermodynamic analysis for Dyson spheres and radiation has also been debated, for instance in \citet{Badescu2014}.  

One important suggestion for the role of a Dyson sphere is that it would be used for computation \citep[e.g.][]{Scharf2023}. Robert J. Bradbury wrote an influential discussion of ``Matrioshka Brains,'' giant machines that perform as many computations as possible from astronomical sources of power like stars.\footnote{\url{https://web.archive.org/web/https://gwern.net/doc/ai/1999-bradbury-matrioshkabrains.pdf}} In that and other works it is imagined that Dyson spheres will have a nested structure (thus their namesake, nested Russian matrioshka dolls) in which outer layers exploit the waste heat of inner layers to optimize computational efficiency.

\subsection{Purpose and Plan of this Paper}

Here, our purpose is to follow Dyson's original philosophy by identifying only the outer limits of what a Dyson sphere might accomplish, and clarifying the appropriate expressions for their efficiency.  We will explore what we can say about the optimal configurations of mass around a star for exploiting energy use for different broad purposes.  Real Dyson Spheres, if they exist, will presumably be both sub-optimal and subject to many practical or other constraints we both can and cannot imagine: our purpose is not to guess those, but to see what we might be able to conclude for Dyson spheres generally.

One of the keys to the analysis in this paper will be the recognition of \citet{Buddhiraju2018}, \citet{Li2020}, and others that the use of optical circulators can, in principle, circumvent some practical limits of heat engines using radiation as a pump and sink. Following the spirit of Dyson, we will accept this as an ``existence proof'' of  technologies that simplify the optimal thermodynamics of Dyson spheres, yielding expressions consistent with the Carnot limit for work efficiency.

Another contribution of this paper will be to clarify how Dyson sphere efficiencies depend on the kind of activities done with stellar radiation, and to explore three general categories. Most previous papers have assumed ``work'' is of the traditional sort, i.e.\ that it is essentially mechanical and is removed from the energy budget of the sphere.  But \citet{Wright2020_Dyson} pointed out this is inappropriate for most work we might imagine a Dyson sphere would do, because Dyson spheres must conserve energy and radiate away \textit{all} of the energy they receive, not just their waste heat.

Finally, this work will find that within a factor of a few, gross Dyson sphere properties are largely insensitive to these details.  This is good: it means that we need not guess too precisely about the means and motives of large technological species to search for them.

\section{Preliminaries}

We begin with a somewhat didactic presentation of the principles of Dyson sphere thermodynamics to orient the reader with the philosophy of the paper.

\subsection{Work, Entropy, and Efficiency}

One typically calculates the amount of work that can be extracted from a source of heat by looking at the energy flow one can generate between a source of heat and a cooler environment / heat sink as heat flows from warm to cold.  One wishes to tap this energy flow to do ``work,'' by which we mean we \textit{remove the energy from the system} and use it to do some sort of (usually mechanical) task. 

This extraction of work removes energy from the system, but not entropy. Since we cannot destroy entropy (we can only increase it), it is impossible to extract all of the energy flow for work---at least some of it must be reserved to carry away the entropy that came into the system.  When this energy leaves the system it is called ``waste heat.''  

The entropy $\delta S$ associated with a small amount of waste heat $\delta Q$ is $\delta Q/T$, meaning that heat at a low temperature has more entropy associated with it per erg than warm temperature heat.  If one expels energy as heat at a lower temperature than one accepted it from the warm source, it will contain more entropy per erg than it had going in. 

We can then heuristically think of a heat engine as something that reallocates entropy across a pool of incoming energy.  The engine puts some energy out as heat into the cooler environment---at this lower temperature, that energy can ``hold'' more entropy than it had coming in, so it can remove most or all of the incoming entropy, freeing up the remaining energy to be extracted as work.

The amount of energy one can extract from a system do to such work is called the \textit{exergy} of the system.  It depends on the amount of entropy the incoming heat started with, and the temperature at which one can expel heat. The maximum efficiency of a machine is the fraction of incoming energy from the hot source that is exergy.

A Carnot engine is an ideal realization of a machine that performs this task optimally, generating no entropy of its own, and has an efficiency given by
\begin{equation}
    \eta_\mathrm{Carnot} = \frac{\dot{W}}{\Edotin} = 1 - \frac{T_h}{T_c}
\end{equation}
\noindent where $\dot{W}$ is the rate at which work is extracted (i.e.\ the rate at which energy, but not entropy, leaves the system) and \Edotin\ is the total energy into the system from the hot source. 

As typically defined, there are many practical difficulties in constructing a heat engine that can achieve this limit using heat transfer, including the need for an infinite heat source and heat sink that do not change their temperature as the machine operates.  It also assumes that one can bring a component of the machine to exactly the temperature of the source or sink---in practice this is difficult or impossible to achieve because heat transfer takes finite time. The Carnot limit is thus technically only achieved for infinitessimal temperature differences, and the work is extracted infinitely slowly. 

For heat engines using a gas or fluid with finite heat capacity exchanging heat with infinite thermal baths acting as pumps and sinks, the ultimate efficiency limit at maximum power is the \textit{endoreversible} limit \citep{Curzon1975}:
\begin{equation}
    \eta_\mathrm{endoreversible} = 1 - \sqrt{\frac{T_c}{T_h}}
    \label{Eq:endoreversible}
\end{equation}
This limit, somewhat surprisingly, is independent of the details of how quickly heat transfer can take place.

A commonly seen expression for thermodynamic efficiency in the limit where radiation at temperature $T_h$ is being used as the heat source and heat is radiated away as radiation with temperature $T_c$, is that of \citet{Petela1964}:
\begin{equation}
    \eta_\mathrm{radiation} = 1 - \frac43 \frac{T_c}{T_h} + \frac13\left(\frac{T_c}{T_h}\right)^4
    \label{Eq:radeff}
\end{equation}
\noindent This form \citep[equation 3.4 in][and repeated in other references]{Landsberg80} assumes that the radiation is isotropic---that is, it assumes the flux on the absorber is $\sigma T_h^4$ and the flux out of the radiator is $\sigma T_c^4$. It is thus inappropriate for situations where the incoming radiation is comes from a small area on the sky, as with solar panels. It is also inappropriate for situations in which energy cannot be exchanged with the environment, since the radiative fluxes in and out do not balance against the work done (i.e.\ it does not assume $\dot{Q}=0$ in the Landsberg formalism below). 

\subsection{Thermodynamics of Radiation}

\label{sec:radiation}
When the source and sink of energy in the system is blackbody radiation, not an infinite-heat-capacity thermal bath at fixed $T$, we must account for the entropy of that radiation with care. \citet{Badescu2014} and \citet{Buddhiraju2018} both present treatments of this case we can draw from.

For radiation, the entropy per time per unit surface area escaping from a blackbody is given by\footnote{Some sources erroneously claim the entropy flux of blackbody radiation is simply $\sigma T^3$, but this ignores the fact that radiant energy has an inherent pressure that adds an extra factor of 4/3.  See \citet{Wu2010}.}
\begin{equation}
\dot{s} = \frac43 \sigma T^3
\end{equation}
\noindent and the energy flux is 
\begin{equation}
F = \sigma T^4
\end{equation}

Significantly, the ratio $\dot{s}T/F = 4/3$, a coefficient that is absent from treatments of efficiencies with respect to heat exchange with thermal baths.  This will have important consequences for our derived limits.

\subsection{The Landsberg Limit and Formalism}

In a foundational paper, \citet{Landsberg80} described the absolute limits of thermodynamic energy conversion of radiation, now referred to as the Landsberg limit.  In their formalism, a machine converts energy with two inputs and four outputs:
\begin{itemize}
    \item Energy flux into a system, $\dot{E}_p$, where $p$ stands for ``pump.'' We will refer to this quantity as \Edotin.
    \item Entropy flux into a system, $\dot{S}_p$ We will refer to this quantity as \Sdotin.
    \item Energy flux into a sink, $\dot{E}_s$. We will refer to this quantity as \Edotout.
    \item Entropy flux into a sink, $\dot{S}_s$.  We will refer to this quantity as \Sdotout.
    \item $\dot{Q}$ the heat flux out of the system into the ambient environment.
    \item $\dot{W}$ the rate of work done by the system, that is, rate of energy out of the system with no corresponding entropy flux.
\end{itemize}

The machine itself has four properties:
\begin{itemize}
    \item $\dot{E}$ the increase in internal energy with time. \item $\dot{S}$ the increase in internal entropy with time.
    \item $T$ the temperature on the boundaries of the machine.
    \item $\dot{S}_g$ the rate of entropy generated by the machine.
\end{itemize}

Global energy conservation is enforced by
\begin{equation}
    \Edotout = \Edotin - \dot{E} - \dot{Q} - \dot{W}
\end{equation}
\noindent and entropy as accounted for with the identity
\begin{equation}
    \Sdotout = \Sdotin - \dot{S} - \dot{Q}/T + \dot{S}_g
\end{equation}

\begin{figure}
    \centering
    \includegraphics[width=\textwidth]{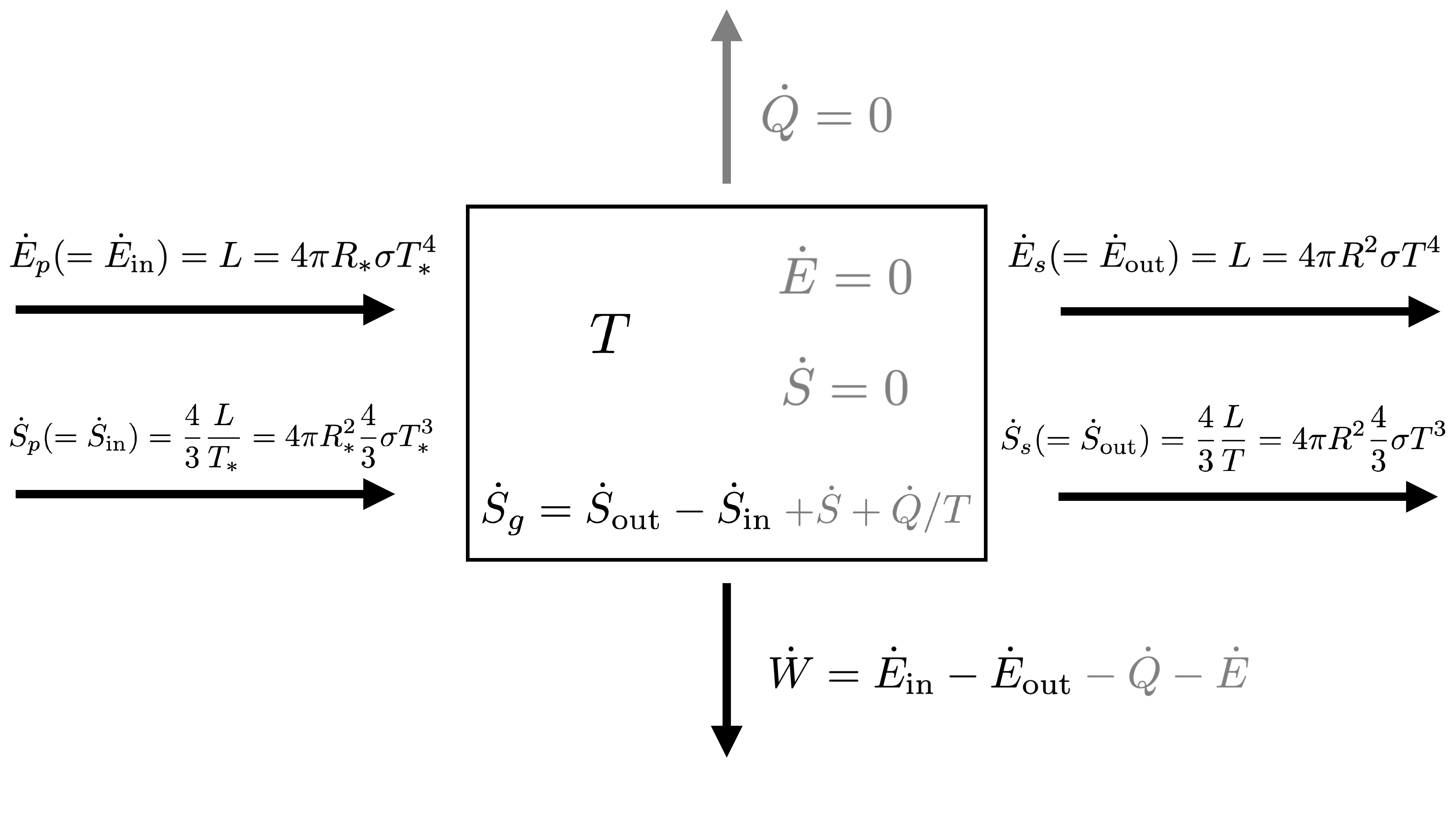}
    \caption{Schematic after Figure 1 of \citet{Landsberg80} showing the inputs and outputs of energy and entropy in a machine.  In that work, inputs and outputs included a ``pump'' and a ``sink'';  we use the terms ``in'' and ``out''. In this work, we apply this formalism to a shell or swarm of material at a distance $R$ from a star. Terms not used in this work because they are zero for Dyson spheres in steady state are in gray. $L$ is implicitly the luminosity of the star, but in systems with feedback or considering incoming radiation from space it may be higher than this. $\dot{Q}=0$ because there is no ambient environment to share heat with except via radiation, already accounted for in \Edotout.  $\dot{S}_g$ will be nonzero for Dyson spheres doing computation, dissipative activities, or working below the Landsberg limit, and $T$ is its outgoing radiation temperature. $\dot{W}$ will be zero unless the sphere is emitting low-entropy radiation, as with a radio beacon. In steady-state, $\Edotin=\Edotout+\dot{W}=L$, $\dot{E}=0$, and $\dot{S} = 0$.  All similar figures in this work are special cases of this master figure, in some cases with $\dot{E}_\mathrm{out,1} + \dot{E}_\mathrm{out,2} = \Edotout$ and $\dot{S}_\mathrm{out,1} + \dot{S}_\mathrm{out,2}= \Sdotout$.
    \label{fig:Landsberg}}
\end{figure}

These are shown in Figure~\ref{fig:Landsberg}.  This figure is a general case of all similar figures later in this work.

Note that the Landsberg formalism is a generalization of calculations that produce the Carnot and radiation limits above, not a contradiction of them.

\subsection{Optical Circulators and the Landsberg Limit}

\label{sec:circulators}

The practical limits of power extraction from radiation are not clear. Many treatments begin with the premise that the incoming radiation must warm an intermediate blackbody to power a Carnot engine, which creates an inefficiency.  In addition, typically when a surface absorbs radiation and comes to some temperature it also emits radiation back towards the source, and a radiator not only disposes of waste heat but absorbs radiation from its environment.  Both effects limit the ability to re-use waste heat at a lower temperature, and to efficiently cool. These effects provide a practical limit, typically below the Carnot limit, for many engine designs using radiation.  

\citet{Buddhiraju2018} (Figure 6) and \citet{Li2020} (Figure 4h) describe schemes using optical circulators to operate at the full Carnot efficiency between two radiation sources that avoids these inefficiencies. The idea is that circulators violate Lorentz reciprocity, which applies to most materials and says that the propagation of light through a system is symmetric with respect to direction of travel.  A circulator exploits polarization to differentiate between the two directions.  In a three port circulator, if light enters port A and exits port B, light entering port B will not exit port A as one would expect under Lorentz reciprocity, but instead exit via port C (and light entering port C emerges from port A). This allows the system to collect light via port A without necessarily returning light via port A, overcoming many practical limitations of heat engines and allowing one to approach the Landsberg limit. 

In their scheme, a set of circulators in series each feed a set of Carnot engines working at lower and lower temperatures.  The first engine operates at $T_h$, the temperature of the incoming radiation. The waste heat from the engine is then passed to another at slightly lower temperature, and another and so on, until the lowest possible temperature for the system is reached.  For their system perfectly coupled to the cold radiation sink, this is the cold temperature $T_l$. By increasing the number of steps between the hottest and coldest temperatures, one reduces the inefficiencies that come from extracting work at finite temperature differences and, in the limit of a very large number of steps, one achieves the Landsberg limit.  

We should also keep in mind that when working with photons as an energy source instead of heat, we might not need to first convert photons to thermal energy and then to electricity or other forms of work; other possibilities may emerge depending on the kind of work to be done. For instance, in photosynthesis photons interact directly with the molecules responsible for plant metabolism; similarly photons might directly perform the activities in question, and thus avoid practical limitations of real heat engines. 

In what follows, we will therefore adopt the schemes of \citet{Buddhiraju2018} and \citet{Li2020} as proofs of principle that we can ignore issues like feedback and intermediate absorbers when computing limits of Dyson spheres.  We will check the validity of this assumption by generalizing our treatment to include other efficiency laws, including the endoreversible case.

\section{Application to Dyson Spheres}

\subsection{Satellites and Dyson Spheres}

Our purpose is to apply the Landsberg formalism properly to the problem of a Dyson sphere.  Dyson spheres have some special properties that make many calculations for work done via radiation in the literature inappropriate.  Four of special relevance are:

\begin{itemize}
    \item Dyson spheres that extract work and perform it locally, for instance if they are doing computation or maintaining biospheres, must eventually radiate their energy away, raising the temperature of their radiators, and resulting in no work extraction $\dot{W}$. This complicates our efforts to define and calculate an efficiency for them.
    \item Dyson spheres use radiation for their input energy and output waste heat.  We thus need to use the fluxes given in Section~\ref{sec:radiation}.
    \item Dyson spheres accept radiation coming from a finite source, i.e.\ from a narrow range of solid angle, and not isotropicaly.\footnote{This is different from a ``dilute'' source of radiation, where blackbody radiation in a given solid angle is at a lower intensity than given by the Planck law by a frequency-independent factor. In both cases less light arrives at a surface, but for different reasons: for ``dilute'' radiation, it arrives from all directions but has been attenuated, as by a gray absorbing medium, and for the case this paper in concerned with it arrives at the full Planck intensity but over a limited range of solid angle.  The thermodynamic difference is in the amount of entropy contained in the radiation: Dilute radiation distributes fewer photons in the same number of modes (directions and frequencies), and since the specific entropy is calculated from the number of ways $M$ photons can occupy $N$ modes via the ratio $M/N$, reducing $M$ with fixed $N$ results in a different specific entropy. For radiation restricted to a solid angle $\Omega$, both the number of photons and the number of modes are reduced by $\Omega/4\pi$, so the ratio $M/N$ is fixed, and the blackbody expression for specific entropy holds. See \citet{Landsberg1979,Wu2010} for a discussion of how to treat dilute radiation.}
    \item The Dyson sphere-star system must conserve energy flux globally. In particular, Dyson spheres can only dispose of energy outwards, not back onto the star or inwards to their own interiors.  Energy sent back to the star will have implications for stellar feedback and potentially the entropy management of the sphere, but will to first order not affect the global energy budget of the system.
\end{itemize}

In this work, we will presume that Dyson spheres are composed of a large swarm of satellites, each presenting a flat ``solar panel'' towards the star and a radiator away from it. We make no assumptions about the nature of the ``solar panel'' except that it collects stellar radiation to perform some sorts of activities, and shares a cross-sectional area with the radiator.  We assume the radiator emits radiation as a blackbody of temperature $T$.

We will model this swarm as being in a thin sphere around the star at radius $R$ and area $4\pi R^2$, as if it were a monolithic shell, which would not be stable \citep{Wright2020_Dyson}. Later we will check to see how moving to a more realistic swarm model changes our answers.

We will assume for this work that the star radiates as a blackbody with temperature $T_*$ and that it has radius $R_*$, and luminosity $L=4\pi R_*^2\sigma T_*^4$. Later, when considering feedback onto the star we will briefly distinguish between the effective temperature $\teff$ defined by the luminosity and radiation, and the blackbody temperature of the surface $T_*$.

Dyson spheres cannot accumulate significant amounts of energy.  As \citet{Wright2020_Dyson} discusses, the luminosities involved are so large that the internal energy of a Dyson sphere that attempts to store energy will quickly become extremely and impossibly high, causing them to become chemically and gravitationally unbound.  If Dyson spheres are long-lived, they must be in steady state, and radiate away all of the energy they collect.

We will thus assume steady state, and that energy is conserved globally.  In terms of the Landsberg formalism, this means we can write:
\begin{eqnarray}
    \dot{Q} &=& 0\\
    \dot{E} &=& 0\\
    \dot{S} &=& 0\\
    \Edotin &=& \Edotout + \dot{W} 
\end{eqnarray}

$\dot{Q}$ is zero because no heat can be transferred to the ambient environment because there is no ambient environment except deep space, and we already account for the transfer of energy and entropy to deep space via \Edotout\ and \Sdotout.  

Note that the radiation from the star is from is given by the blackbody law, but it only arrives from a small range of solid angles when it gets to the sphere.  We account for this by interpreting terms like $\dot{E}$ to have units of power and not flux---that is, we explicitly account for the area of our collectors and radiators.  Note that this differs from many treatments, including \citet{Landsberg80}, which somewhat confusingly and tacitly ignore this distinction, a choice that is equivalent to assuming all radiation fields are isotropic (i.e.\ incoming radiation arrives at a surface uniformly in all directions).

\subsection{Circulators for Dyson Spheres}

We next need to consider all of the radiation into and out of the Dyson sphere from both directions.  For simplicity we will ignore in this work the effects of radiation from deep space and presume it is effectively zero, but in practice for very large spheres it sets an outer limit to the size and efficiency of a sphere.

More important will be radiation from the inside surface of a sphere. This radiation will mostly land on the inside of the sphere that radiated it, for no net effect, but some will be returned to the star or an inner sphere. This will alter the energy balance of the star and sphere, and some treatments account for this feedback explicitly \citep[e.g.][]{Wright2020_Dyson,Huston2022}.

However, since we are looking at the outer limits of what is possible, we can invoke circulators to ignore this effect, as described in Section~\ref{sec:circulators}.  Consider that the shell makes use of three-port circulators, with Port I pointing inward, Port O pointing outward, and Port E connecting to the work extractor (and engine). The circulator can circulate radiation in the cycle $O\rightarrow I \rightarrow E \rightarrow O$.  Doing so would mean each shell would ``see,'' when looking outward, the interstellar radiation field, which it would pass along inwards without absorbing it, and would accept radiation from the inner shell moving outward without returning any radiation from its own emissions. 

\begin{figure}[!htb]
    \centering
    \includegraphics[width=\textwidth]{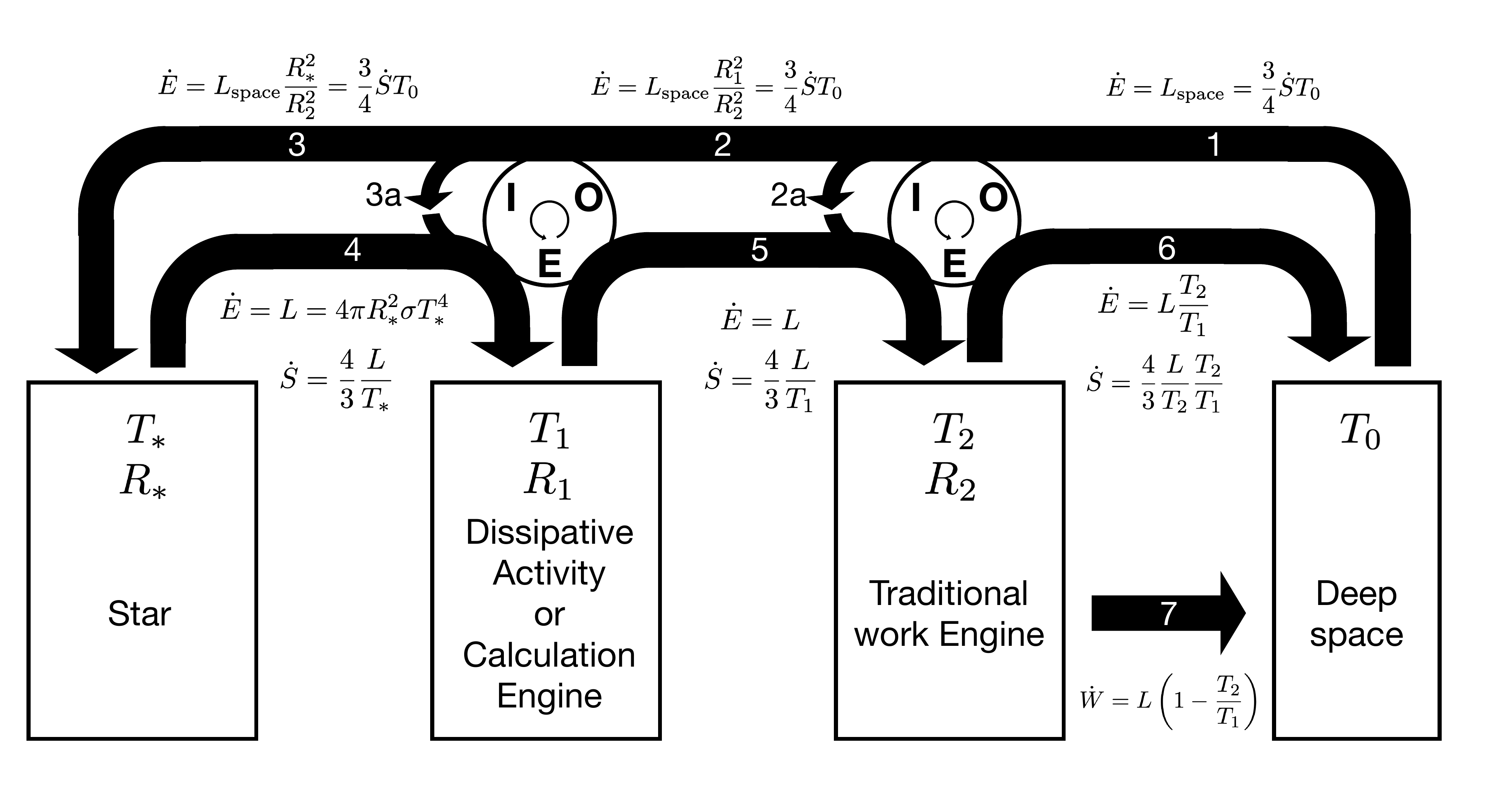}
    \caption{Schematic illustrating how to control the flow of radiant energy in a system with one or more Dyson spheres such that there is no radiative feedback inwards.  Circulators have three ports, pointing inward (``I''), outward (``O''), and towards the work extractor / engines in each sphere (``E''). Power from deep space $L_\mathrm{space}$ is passed inward by each sphere, striking either the outside of the next smallest sphere (at the next port ''O'', paths 2 and 3) or the interior surface of the sphere that passed it on (at its own port ``I'', paths 2a and 3a). Figure~\ref{fig:geometry} shows the 2D geometry corresponding to this scheme.  We have ignored the small contribution of $L_\mathrm{space}$ and its entropy to the function of the engines except to show its path via small arrows for completeness, but in principle it adds to the outwards energy flow. Here we show a specific worked example involving two nested spheres doing different activities, but the scheme generalizes.  This demonstrates that our simple treatment in this paper that ignores ingoing radiation from the spheres is a physically possible limit.
    \label{fig:Circulators}}
\end{figure}

\begin{figure}[!hbt]
    \centering
    \includegraphics[width=\textwidth]{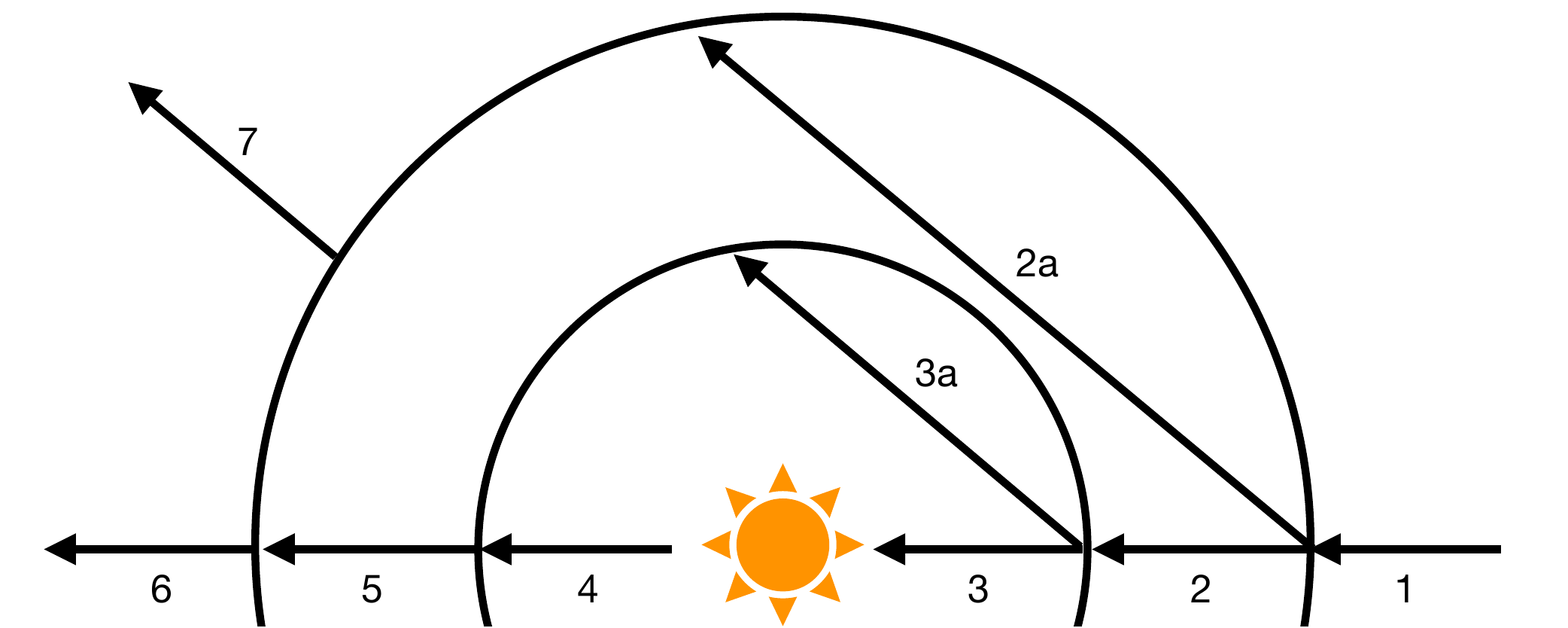}
    \caption{Schematic showing the correspondence between the numbered paths of flow in Figure~\ref{fig:Circulators} and the geometry of the Dyson sphere.}
    \label{fig:geometry}
\end{figure}

Figure~\ref{fig:Circulators} and \ref{fig:geometry} show the scheme as a worked example of two nested spheres doing different kinds of activities, as described in the next section.

Putting this together with Section~\ref{sec:radiation}, and ignoring the very small amount power and entropy from deep space, we therefore have for a single Dyson sphere around a star

\begin{eqnarray}
\Edotin &=& 4\pi R_*^2 \sigma T_*^4\\
\Sdotin &=& 4\pi R_*^2 \frac43 \sigma T_*^3\\
\Edotout &=& 4\pi R^2 \sigma T^4\\
\Sdotout &=& 4\pi R^2 \frac43 \sigma T^3
\end{eqnarray}

All of these are shown in Figure~\ref{fig:Landsberg}.

Figure~\ref{fig:Circulators} has a superficial similarity to the schemes of \citet{Buddhiraju2018} (Figure 6) and \citet{Li2020} for achieving the Landsberg limit with radiation (because it was inspired by them), but it is not identical. First, if we have nested shells they will generally be widely spaced, with large temperature differences among them. Secondly, we must account explicitly for the geometry of spheres and very different radiating areas of our components, specifically with respect to the path of light from deep space, as shown in Figure~\ref{fig:geometry}.  Finally, because we have finite radiating area, we cannot operate at the Carnot efficiency against the cold radiation of space (i.e.\ $T_0$ in Figure~\ref{fig:Circulators}), but only against the radiation temperature of our sphere.

\subsection{Three Kinds of Dyson Sphere Stellar Engine Activities}

In order to make use of this formalism, we need to define what we mean by work and how we wish to quantify the activities the Dyson sphere can do.  There are three options we will consider:

\begin{enumerate}

\item \textbf{Computation:} We can imagine the sphere extracts electrical energy from starlight, uses it to run a computer to do a calculation, and then this computer disposes of the energy as heat. We would thus measure the efficiency of such a computer not based on the energy it consumes, but on the rate of computations it can perform, $r$. (Our analysis will be general, independent of the actual internal mechanism used to power the computer). 

\item \textbf{Dissipative Activities:} Most energy use is dissipative. Consider that nearly all work done by technology on Earth---the majority of which is not computation---ultimately dissipates into heat. Fossil fuels used for transportation are temporarily converted to kinetic energy, but all machines suffer friction which converts that energy to heat. This is also true of biological activity: plants convert solar energy into chemical energy, animals consume plants and use this energy in their metabolisms, and in all cases all of this energy is ultimately carried away as heat into the environment.  In this case, which is a generalization of computation, there will be some sort of internal rate of power put to use, \Winternal, before it cascades into waste heat.  Here, ``internal'' means it is internal \textit{to the Dyson sphere}; we will conceptually have to define part of the sphere to be the work extractor, from which work \Winternal\ leaves, even if work never leaves the sphere itself.  

The distinction between cases 1 and 2 is purely in terms of how we will measure the efficiency of the sphere. In both cases, since no work leaves the sphere, with respect to Figure~\ref{fig:Landsberg} we will write
\begin{equation}
\dot{W} = 0    
\end{equation}

\item \textbf{Traditional Work:} Dyson spheres might convert starlight into a low entropy form that leaves the sphere, for instance as a strong, coherent radio signal.  This case will track the usual calculations for thermodynamic efficiency best, since they are usually concerned with the rate of work $\dot{W}$ that leaves the system, which is how we will measure our efficiencies in this third case. In this case, the entire sphere is a work extractor.
\end{enumerate}

\section{Optimal Efficiencies for Dyson Spheres}

\subsection{Computation}

In a computation, each binary logical operation requires the creation of $S=k\ln{(2)}$ of entropy (from erasure of memory in the system).\footnote{Reversible computing \textit{might} be able to overcome this limit, but even if a Dyson sphere computer performs such calculations, it will still need to overcome some rate of error generation, generating entropy equal to $k\ln{(2)}$ per bit.  Either way, the computer is limited in its computational ability by its ability to dispose of this entropy.} The Landauer limit is derived from this fact, stating that one must expel at least $TS = kT\ln{(2)}$ of energy as heat after performing a binary logical operation. 

An ideal extractor of computation from light would work something like this.  A machine receives a very small amount of radiant power $\delta \dot{E}$ and does $\delta \dot{E} / (kT \ln{(2)})$ calculations per unit time with it, generating $\delta \dot{E}$ in waste heat with $\delta \dot{S} = \delta \dot{E} / T$ entropy per unit time. 

This heat is radiated away, causing its radiator to increase its temperature such that its radiant power output $P = A \sigma T^4$ increases by $\delta \dot{E}$, where $A$ is its area.  Differentially, we have $\delta P = 4 A \sigma T^3 \delta T = \delta \dot{E}$.

We can then integrate the total effect of an arbitrary, macroscopic amount of energy flux $\dot{E}$, starting with a very cold radiator.  The total energy flux through the system is then
\begin{eqnarray}
    \Edotout &=& \int_0^{\dot{E}}\delta\dot{E} \\
    &=& \int_0^T 4 A \sigma T^3 \delta T \\
    &=& A \sigma T^4
\end{eqnarray}
\noindent as it must be, and the total entropy emitted is
\begin{eqnarray}
    \Sdotout &=& \int_0^{\dot{E}} \frac{\delta\dot{E}}{T} \\
    &=& \int_0^T 4 A \sigma T^2 \delta T \\
    &=& \frac43 A \sigma T^3
\end{eqnarray}
as it must be. The total number of calculations done in this case is 
\begin{equation}
    r = \frac{\Sdotout}{k\ln{(2)}} = \frac43 \frac{L}{kT\ln{(2)}}
\end{equation} 
\noindent where $L$ is the luminosity of the star.

The factor of $\frac43$  is not a violation of the Landauer limit but arises from the fact that the Landauer limit comes from the thermodynamic definition of heat and entropy, $dQ = TdS$, which is a \textit{differential} relationship---it does not imply that $Q/S = T$ for macroscopic amounts of heat, so we need not necessarily spend $kT\ln{(2)}$ per calculation in bulk. We can think of some operations generating heat that is radiated away in the low temperature modes of the blackbody radiation field, having a lower effective $T$ than the surface of the sphere. 

\begin{figure}[!hbt]
    \centering
    \includegraphics[width=\textwidth]{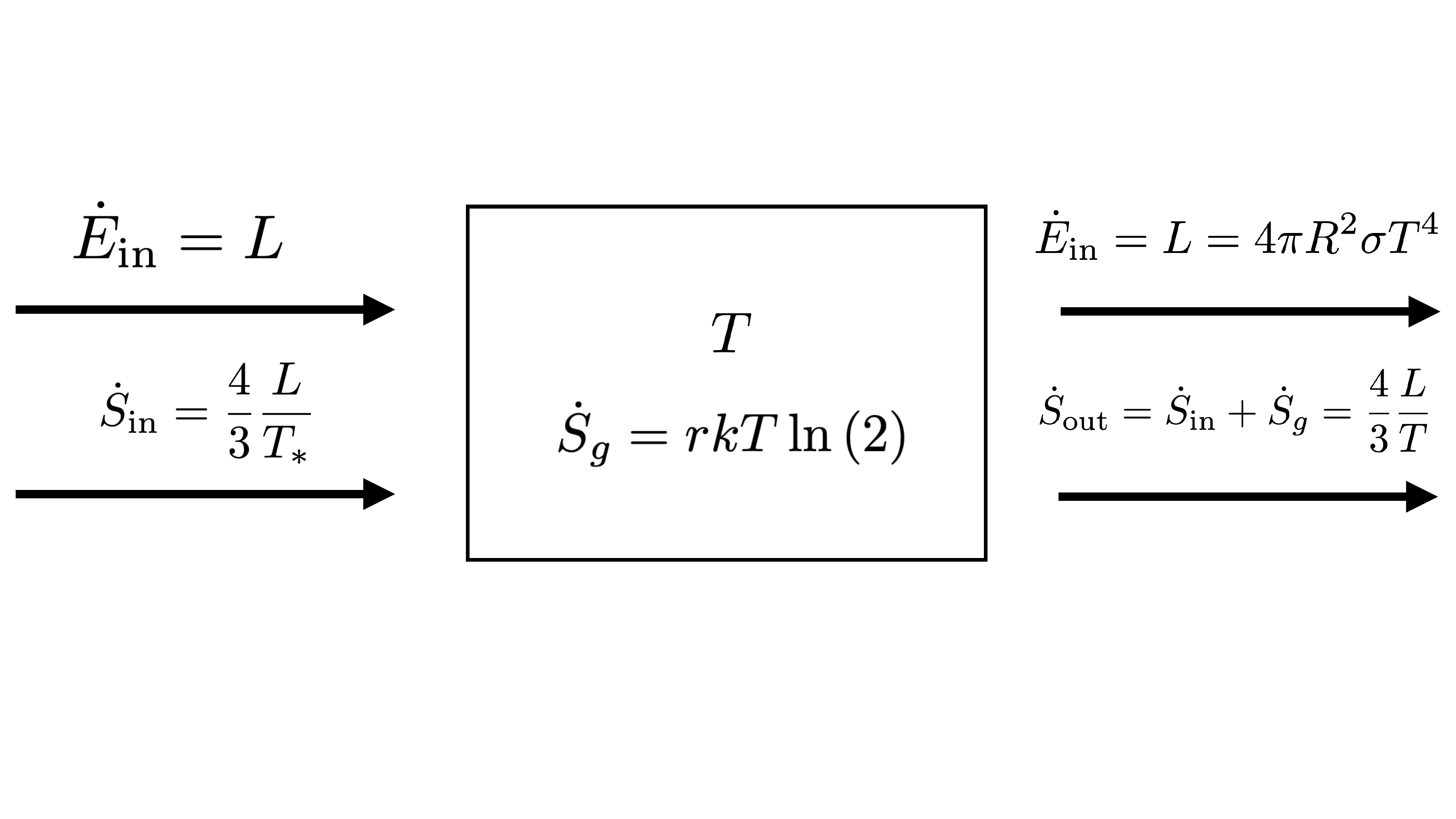}
    \caption{Schematic for a Dyson sphere doing pure computation. $\dot{W}=0$ and is not shown because all incoming energy is used for computation and expelled as waste heat. At maximum efficiency, all of the entropy added to the incoming energy is generated by computation at rate $r$ at the Landauer limit, giving rise to an internal entropy generation rate $\dot{S}_g$.  $T$ is determined from energy balance, given $R$.  From this, $\dot{S}_g$ can be computed from entropy balance, yielding $r$.}
    \label{fig:computation}
\end{figure}

We can now apply Landsberg's formalism properly to the case of a Dyson sphere performing calculations as illustrated in Figure~\ref{fig:computation}.

By energy balance, because $\dot{W}=0$ we have
\begin{equation}
    \Edotin = L = 4\pi R_*^2\sigma T_*^4 = \Edotout = 4\pi R^2\sigma T^4
\end{equation}
\noindent Since we consider $L$ and $R$ to be given, this uniquely determines $T$:\footnote{We could also fix $T$ to determine $R$, of course.} 

\begin{equation}
T = T_*\sqrt{\frac{R_*}{R}}
\end{equation}

With $T$ determined, we can then examine entropy balance to determine $\dot{S}_g$:
\begin{eqnarray}
    \dot{S}_g = \Sdotout - \Sdotin
\end{eqnarray}
\noindent yielding
\begin{eqnarray}
    r &=& \frac43 \frac{L}{kT\ln{(2})} \left[1-\frac{T}{T_*}\right] \label{eq:compT}\\
    &=& \frac43 \frac{L}{kT_*\ln{(2})} \left[\sqrt{\frac{R}{R_*}}-1\right]\label{eq:compR}
\end{eqnarray}

The factor in square brackets in Eq.\ref{eq:compT} is the Carnot efficiency, so we may \textit{heuristically} say that the sphere extracts work at the Carnot efficiency and uses it do to calculations at a cost of $\frac34 kT \ln{(2)}$ per calculation, appropriate for radiation.  Note that the low temperature in both these calculations is \textit{not} the ambient radiation around the sphere (i.e.\ the interstellar radiation field, whose small contribution we have ignored), but the surface temperature of the machine.  This is enforced by energy conservation, the steady state assumption, and the area of the sphere.

\subsection{Dissipative Activities}

\begin{figure}[!hbt]
    \centering
    \includegraphics[width=\textwidth]{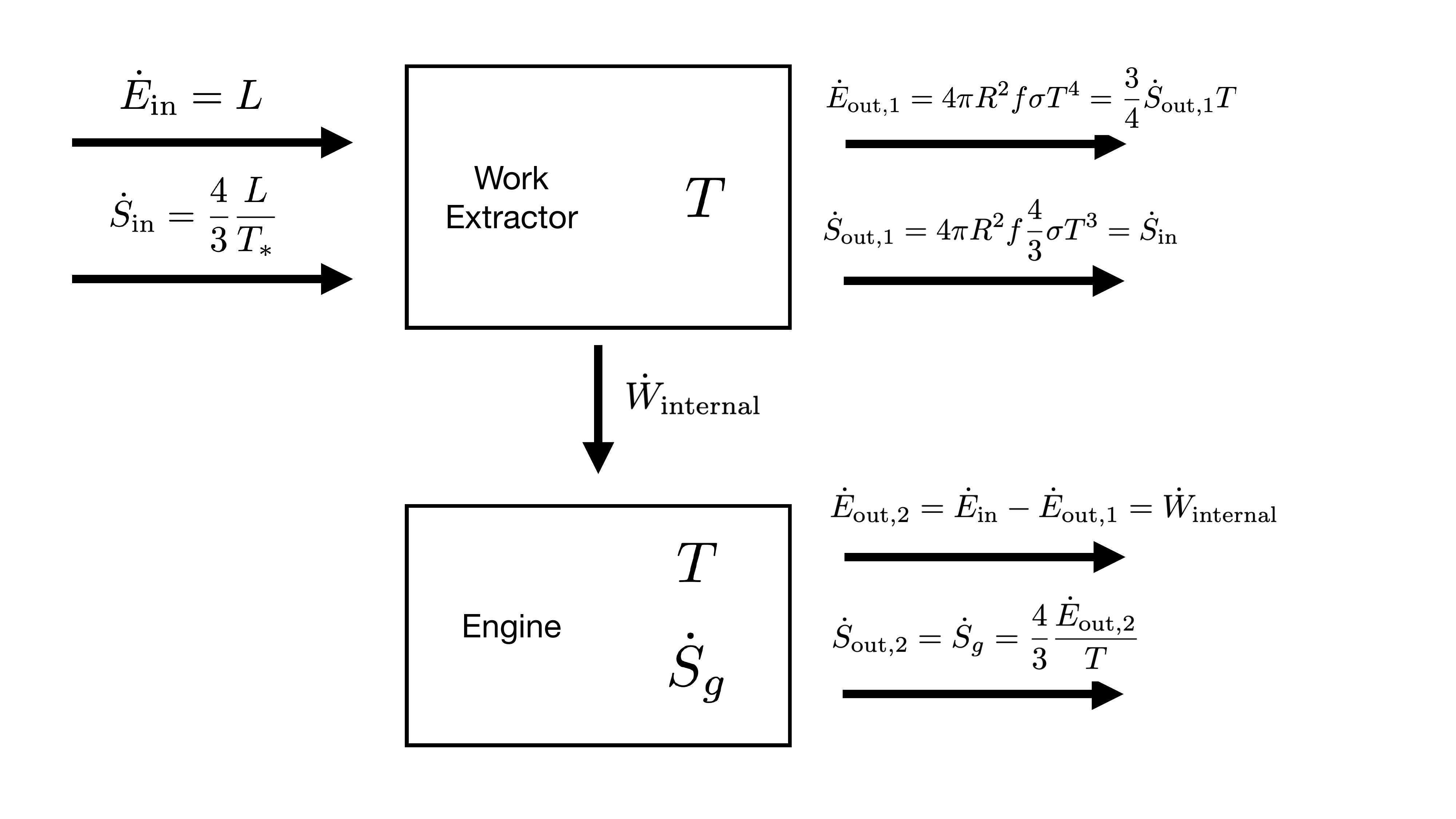}
    \caption{Schematic showing the notional two-part machine used to do dissipative mechanical activities.  The energy extractor generates work \Winternal, which is passed on to an ``engine,'' generating entropy at rate $\dot{S}_g$. Importantly, the two systems share radiators (the extractor uses a fraction $f$ of them, the engine the rest) so they share a common temperature.  The two outputs combine to result in the same scheme as shown in Figures~\ref{fig:Landsberg} and \ref{fig:computation}.  Global energy balance determines $T$ via $L=4\pi R^2 \sigma T^4$, and entropy and energy balance together in the work extractor alone determine \Winternal.}
    \label{fig:TwoPart}
\end{figure}

We can determine the amount power a Dyson sphere can devote to dissipative activities by heuristically breaking the system into two pieces: a work extractor generating $\dot{W}_\mathrm{internal}$ of work per unit time in the usual thermodynamic sense, and an ``engine'' that makes use of this, which ultimately dissipates as heat.  We imagine the sphere dedicates only a fraction $f$ of its radiating surface to passing along entropy received from the star as waste heat from the extractor, and then sends the resulting work across its boundary to an engine which eventually radiates the energy away from the engine using the remaining radiator fraction $(1-f)$.  This is shown schematically in Figure~\ref{fig:TwoPart}.  

The efficiency of the sphere now can be expressed as the efficiency of the work extractor $\eta = \dot{W}_\mathrm{internal}/\Edotin$.  Importantly, this scheme is simply a more detailed accounting of energy flow than that in Figure~\ref{fig:Landsberg}, and could also be interpreted as a way to implement Figure~\ref{fig:computation}. An important consequence of this scheme is that the radiators for the work extractor and engine have a common temperature.\footnote{Changing this by, for instance, making $f$ larger amounts to an intermediate case between dissipative activities and traditional activities. For $f=1$, we recover the traditional activity limit described in the next section.}

At maximum efficiency the work extractor produces no entropy, so 
\begin{equation}
\Sdotin = \dot{S}_\mathrm{out,1}
\end{equation}  

We also have from energy balance that 
\begin{equation}
\Edotin = \dot{E}_\mathrm{out,1} + \dot{E}_\mathrm{out,2} = L
\end{equation}
\noindent Together with the values shown in Figure~\ref{fig:TwoPart}, these two equations allow us to solve for $f$ and $T$ 
\begin{eqnarray}
T &=& T_*\sqrt{\frac{R_*}{R}}\\
f &=& \frac{T}{T_*} = \sqrt{\frac{R_*}{R}}
\end{eqnarray}
\noindent and so by simple energy balance we have
\begin{eqnarray}
    \dot{W}_\mathrm{internal} = \Edotin -  \dot{E}_\mathrm{out,1} = (1-f)L
\end{eqnarray}
\noindent yielding the familiar Carnot efficiency
\begin{equation}
\eta = \frac{\dot{W}}{\Edotin} = 1-\frac{T}{T_*} = 1 - \sqrt{\frac{R_*}{R}}
\label{eq:dissLand}
\end{equation}
Using radiation as a source of energy and a sink of waste heat thus allows a work extractor to operate at the Carnot efficiency, consistent with our result for computational rate.  

\subsection{Traditional Work (That Leaves the Sphere)}

In the case where the purpose of the sphere is to do work that leaves the sphere, we have a different relation.  This work might go out as coherent radiation as in a radio beacon, for instance.\footnote{This case would also cover something exotic like energy-to-mass conversion (antimatter might be a worthwhile output product). In this case the energy merely changes form, it does not ``leave'' the sphere, but the essence  is the same: we have a colder sphere, because the  extracted work is not warming the radiators.}

In this case, we have the same schematic as Figure~\ref{fig:TwoPart} except that the ``engine'' is not present, the energy $\dot{W}$ leaves the system entirely, and $f=1$.

Operating at maximum efficiency, we have $\Sdotin=\Sdotout$  We then have for the temperature of the outgoing radiation and efficiency

\begin{eqnarray}
    T &=& T_* \left(\frac{R_*}{R}\right)^\frac23 \label{eq:tradtr}\\
    \dot{W} &=& L \left( 1 - \frac{T}{T_*} \right) \label{eq:tradtt}
\end{eqnarray}
\noindent yielding efficiency
\begin{equation}
\eta = 1 - \frac{T}{T_*} = 1 - \left(\frac{R_*}{R}\right)^\frac23
\label{eq:tradLand}
\end{equation}

That is, it satisfies the Carnot efficiency but at a lower temperature (and, so, higher efficiency) than the dissipative case, going as the $\frac23$ power of $R_*/R$ of the stellar temperature instead of the square root.  

\begin{figure}[!hbt]
    \centering
    \includegraphics[width=\textwidth]{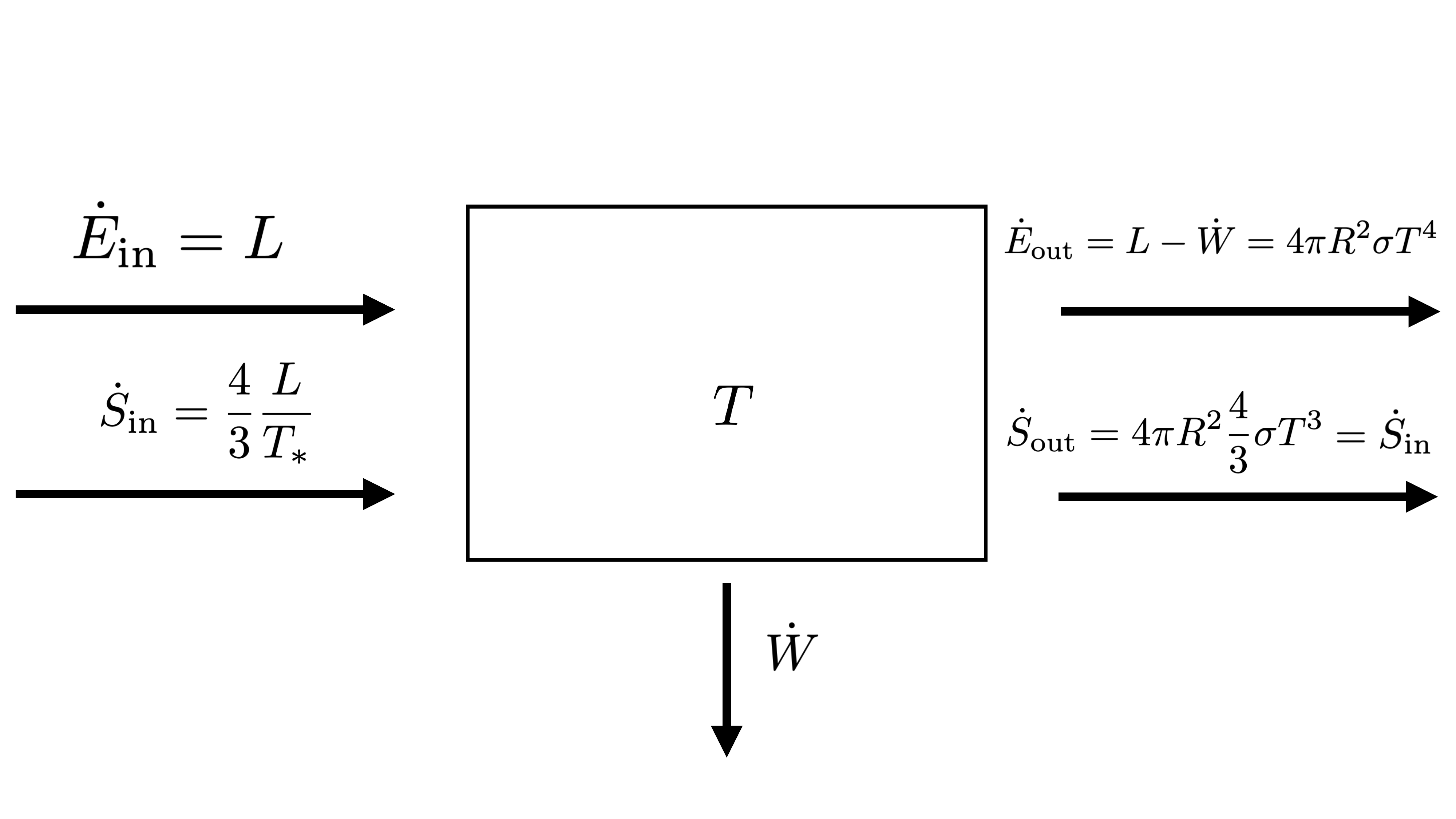}
    \caption{Schematic for work that leaves the sphere, for instance as coherent radio emission (having negligible entropy). In this case, entropy balance determines $T$, from which energy balance determines $\dot{W}$. }
    \label{fig:emission}
\end{figure}

Figure~\ref{fig:emission} illustrates the scheme.

We can now see why the Petela efficiency, Eq.~\ref{Eq:radeff}, does not apply to Dyson spheres.  \citet{Badescu2023} derives a result for this equation in the case where incoming radiation occupies a range of solid angle with geometric factor $f_\mathrm{in}$ and outgoing radiation a has a geometric factor $f_\mathrm{out}$ (where I have simplified the notation):

\begin{equation}
    \eta_\mathrm{radiation} = 1 - \frac43 \frac{T}{T_*} + \frac13\frac{f_\mathrm{in}}{f_\mathrm{out}}\left(\frac{T}{T_*}\right)^4
    \label{eq:Badescu}
\end{equation}

By the definition of the geometric factor $f\equiv \int_\mathrm{source} \cos{\theta} d\Omega$ \citep[][Eq.~16]{Badescu2023}, a sphere of radius $R_*$ overhead at distance $R$ has $f_\mathrm{in} = \pi(R_*/R)^2$, and for our isotropic outward radiation, we have $f_\mathrm{out}=\pi$.  From Eq.~\ref{eq:tradtr}, which enforces $\dot{Q}=0$, we have $f_\mathrm{in} = (T_*/T)^3$, with which substitution Eq.~\ref{eq:Badescu} reduces to the Carnot efficiency, consistent with Eq~\ref{eq:tradtt} and \citet{Badescu2023} Table 7. Eq.~3.  There is thus no contradiction between our results and those of \citet{Petela1964}.

\subsection{Endorevesible Limits}

As an example of a ``practical'' solution to energy extraction for sunlight, we will also imagine that an engine would operate below the Landsberg limit because it needs to convert incoming starlight to heat to be exchanged in a traditional heat engine, which can only be done with some finite rate, yielding an efficiency given by, for instance, Eq.~\ref{Eq:endoreversible}.  Under this concession, we imagine that this conversion generates some entropy that must expelled in addition to the ordinary waste heat and any heat from dissipative activities.

\begin{figure}[!hbt]
    \centering
    \includegraphics[width=\textwidth]{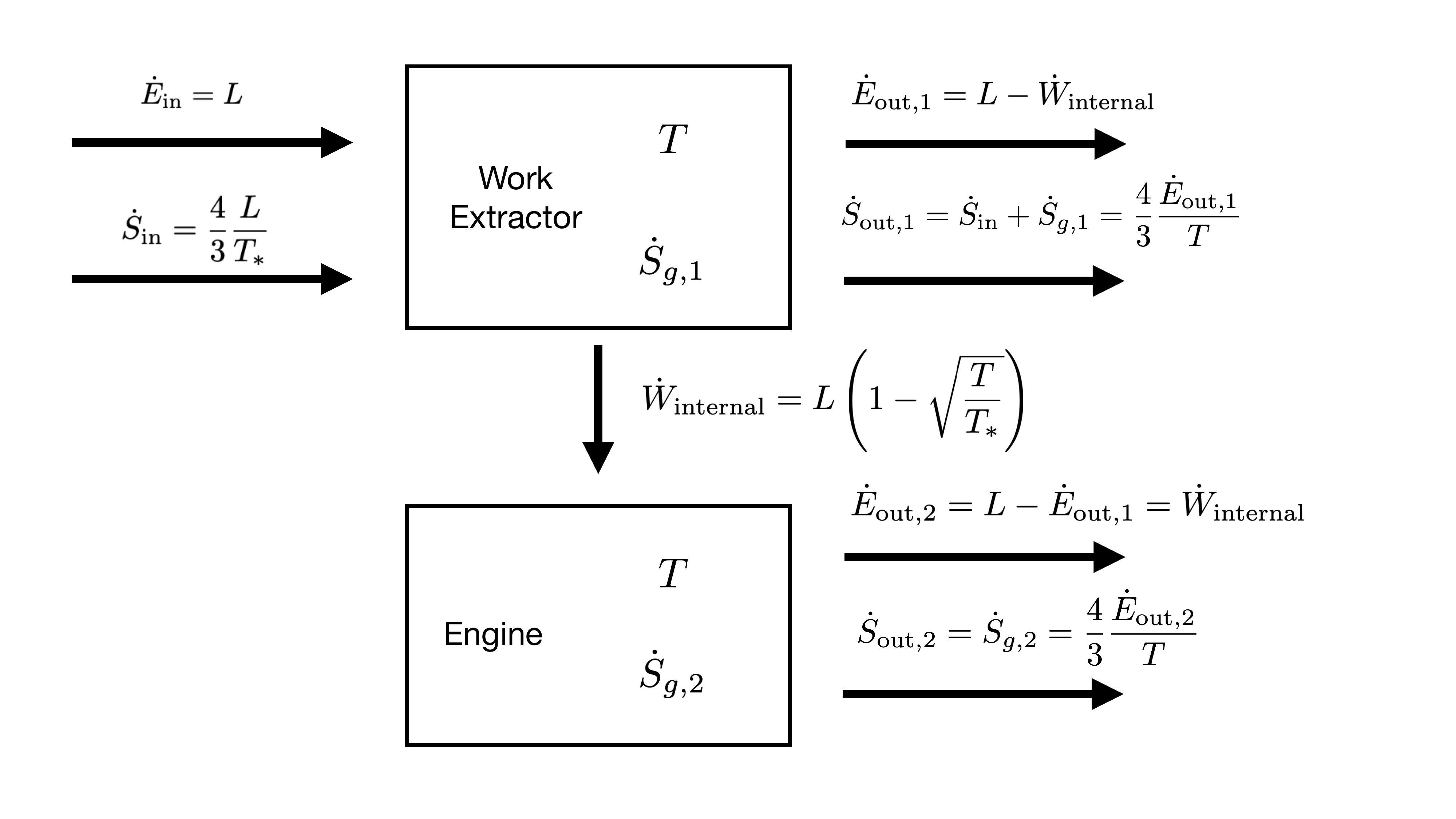}
    \caption{Schematic for endoreversible computation or dissipative activities in a Dyson sphere. In this case, some extra entropy $\dot{S}_{g,1}$ is generated by the work extractor, since it operates below the Carnot limit. Here, energy balance and $R$ determine $T$ as usual via $L = 4\pi R^2\sigma T^4$ for dissipative activities and computation. For computation, the computational rate $r$ is then determined from $\dot{S}_{g,2}$, equal to $\frac43\Winternal/T$. (For traditional work, the engine is not present and we replace \Winternal\ with $\dot{W}$, and then calculate $T$ from $L - \dot{W} = 4\pi R^2\sigma T^4$.)}
    \label{fig:endoreversible}
\end{figure}

We consider the notional two-part, extractor-engine scheme in the endoreversible limit in Figure~\ref{fig:endoreversible}

Here, $T$ has the same values as in the dissipative and calculation cases in the Landsberg limit.  For computation, we have
\begin{eqnarray}
    r &=& \frac{\dot{S}_{g,2}}{k\ln{(2)}}\\
      &=& \frac43 \frac{L}{k\ln{(2)}} \left( \frac{1}{T} - \frac{1}{\sqrt{TT_*}}\right)\\
      &=& \frac43 \frac{L}{k\ln{(2)}} \left( \sqrt{\frac{R}{R_*}} - \left(\frac{R}{R_*}\right)^\frac14 \right)
\end{eqnarray}
For dissipative activities we have 
\begin{equation}
\eta = 1 - \sqrt{\frac{T}{T_*}} = 1 - \left(\frac{R}{R_*}\right)^\frac14 
\label{eq:dissendo}
\end{equation}
\noindent This situation for traditional work is different than in the Landsberg limit, because we have extra entropy generated in the work extractor, $\dot{S}_{g,1}$. Since $\Winternal$ is now specified, we instead determine the temperature via energy balance
\begin{equation}
    \Edotout = L - \Winternal = L \sqrt\frac{T}{T_*}
\end{equation}
\begin{equation}
    4\pi R^2 \sigma T^4 = 4\pi R_*^2 \sigma T_*^4 \sqrt\frac{T}{T_*} \\
\end{equation}
\begin{equation}
T = T_* \left(\frac{R_*}{R}\right)^\frac47
\end{equation}
\noindent yielding the efficiency
\begin{equation}
\eta  = 1 - \sqrt{\frac{T}{T_*}} = 1 - \left(\frac{R}{R_*}\right)^\frac27 
\label{eq:trandendo}
\end{equation}

Fig.~\ref{fig:swarmseff} shows the efficiencies in the four cases (Eqs.~\ref{eq:dissLand},\ref{eq:tradLand},\ref{eq:dissendo},\ref{eq:trandendo}) as a function of total shell area.

\subsection{Swarms of Material}

Real Dyson spheres will likely be composed of swarms of material, potentially at a large range of distances.  The optimal arrangement of these will depend on many details we are unlikely to be able to guess, but we can at least get a sense of how those details matter by bounding the problem.

We will assume the total cross-sectional area of a large number of individual satellites is $A$.  In the ideal case, elements of a swarm will never shadow each other, and as their number increases they will need to make ever more elaborate maneuvers to avoid this. In the limit that they capture all of the light from the star, we would then have $A=4\pi R^2$, but this may be this may be impossible to achieve.

Another bound might be to consider that the swarm elements avoid collisions but do not avoid shadowing, and so regularly block each others' views.  For large numbers of small satellites in random orbits, we might expect this to result in the swarm having optical depth 
\begin{equation}
\tau = \frac{A}{4\pi R^2}
\end{equation}
\noindent where $A$ is the total cross-sectional area of the elements, and $R$ is their typical orbital distance. Note this is an inappropriate approximation for shells, which we assume have $\tau=\infty$.

In this case, the total energy collected $\Edotin$ will be lower than that we would expect for a partial shell of area $A$ by a factor 
\begin{equation}
1-e^{-\tau}
\end{equation}
\noindent as will our rate of computation or work.  If we still compute efficiencies in terms of $L$, this will decrease all of our efficiencies by that factor, as well.

In the limit of a very small amount of mass---say, just a single satellite---this factor becomes
\begin{equation}
    1-\exp{\left(-\frac{A}{4\pi R^2}\right)} \rightarrow \frac{A}{4\pi R^2}
\end{equation}

\subsection{Summary of Results for Dyson Spheres}

Summarizing these results, we can write in general that the efficiency of a Dyson sphere for work is
\begin{equation}
\eta = (1-e^{-\tau}) \left[1-\left(\frac{T}{T_*}\right)^m\right]
\end{equation}

\noindent where $\tau \rightarrow \infty$ for a shell and $\tau = A/4\pi R^2$ for swarms of total cross-sectional area $A$; and where $m=1$ in the Landsberg limit, and $m=1/2$ in the endoreversible limit.  We might imagine that other practical limitations yield some other value for $m$ such that $m<1$. 

Our relation for the temperature is

\begin{equation}
 T = T_* \left(\frac{R_*}{R}\right)^n
\end{equation}

\noindent where $n=\frac12$ for dissipative activities and computation, and $n=2/(4-m)$ for traditional work (so, $\frac23$ in the Landsberg limit, and $\frac47$ in the endoreversible limit.)

Putting this together, the efficiency of a sphere for work in terms of $R$ is 

\begin{equation}
\eta = (1-e^{-\tau}) \left[1-\left(\frac{R_*}{R}\right)^{nm}\right]    
\end{equation}

\noindent where the exponent $nm$ ranges in our examples from $\frac14$ to $\frac23$. 

Finally, the rate of computation is 
\begin{eqnarray}
    r &=& \frac43 \frac{L}{kT\ln{(2})} (1-e^{-\tau}) \left[ 1 - \left(\frac{T}{T_*}\right)^m\right]\\
    &=& \frac43 \frac{L}{kT_*\ln{(2})} (1-e^{-\tau}) \left[ \sqrt{\frac{R}{R_*}} - \left(\frac{R}{R_*}\right)^\frac{1-m}{2}\right] \label{eq:generalrate}
\end{eqnarray}
\noindent where the exponent $(1-m)/2$ is zero in the Landsberg limit and $\frac14$ in the endoreversible limit.

\section{Optimal Configurations}

Next, we turn to see if we can deduce any observational consequences of the various assumptions we have made.  We will especially look for any results that are rather insensitive to our assumptions, as they might be general properties of Dyson spheres.  

\subsection{Absence of Mass Constraints}

In all cases, we see that for a single shell, the most efficient configuration is one that maximizes $R$ and minimizes $T$, performing as much work or computation as possible on the luminosity $L$. The more mass one has at one's disposal, the larger the sphere that can be built.

There are strongly diminishing returns, however. Let us look just at the case of a shell doing dissipative activities in the Landsberg limit. A Dyson sphere at $\about 1$ au around a Sun-like star would have roughly $\eta = 1 - \sqrt{300/6000} = 95\%$ efficiency. This is already very close to unity; no amount of mass or sophistication of technology can improve this by an order of magnitude.

Further, since the mass $M$ required for a sphere presumably scales with its area $A$, we have that the amount of ``wasted'' energy from not increasing the size of a sphere to be colder and more efficient scales as:

\begin{equation}
1 - \eta \propto R^{-\frac12} \propto A^{-\frac14} \propto M^{-\frac14}
\end{equation}

This means that to capture half of the unused energy, one must increase the mass of the sphere by a factor of  $2^4 = 16$.

The situation is not much better for computation.  There, we have for large $R$:
\begin{equation}
r \propto \sqrt{R} \propto A^{\frac14}\propto M^{\frac14}
\end{equation}
So doubling the amount of mass increases the computation rate by a factor of $2^\frac14$ or around 20\%. 

Thus, the efficiency of a sphere measured as work done (or computations performed) \textit{per gram} drops sharply with radius. Since presumably there is some cost to acquiring mass,  with such strongly diminishing returns we might expect Dyson spheres to be reasonably warm, corresponding to efficiencies of 10s of percent.

\subsection{Optimum Shell Radius for Fixed Surface Area (Mass)}

In the opposite limit, we might ask what the optimum distance is for a very small amount of mass, composing, say, a single satellite of area $A$. In this case, we do not have the entire luminosity of the star to work with, and we must balance proximity (which gives us more flux) with efficiency.  In this case the total work done will be 
\begin{equation}
\dot{W} = L\frac{A}{4\pi R^2} \left[1-\left(\frac{R_*}{R}\right)^{nm}\right]
\end{equation}
This function is maximized at $R = R_*\left(\frac{nm+2}{2}\right)^\frac{1}{nm}$, the coefficient of which for our values is around $\about1.6$, so $R$ is extremely close to the star.  

The lesson is that higher fluxes trump higher efficiencies up to very high temperatures---nearly as hot as the star---and so for small amounts of mass the optimal configuration is as close to the star as the engine allows, perhaps at the limits where its components would begin to melt.

As the mass (area) available grows, there is a balance between adding new elements at this minimum distance and avoiding shadowing.  At some point, one will have enough mass that an optically thick shell is formed, and the best use of additional mass involves expanding the shell to lower its optical depth.  We then shift to maximizing the function
\begin{equation}
    \dot{W} = L \left[1-\exp{\left(-\frac{A}{4\pi R^2}\right)}\right] \left[1-\left(\frac{R_*}{R}\right)^{nm}\right]
\end{equation}

\begin{figure}
    \centering
    \includegraphics[width=0.49\textwidth]{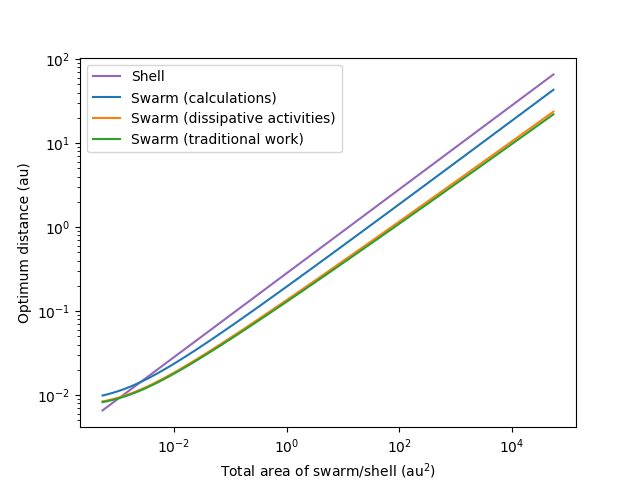}
    \includegraphics[width=0.49\textwidth]{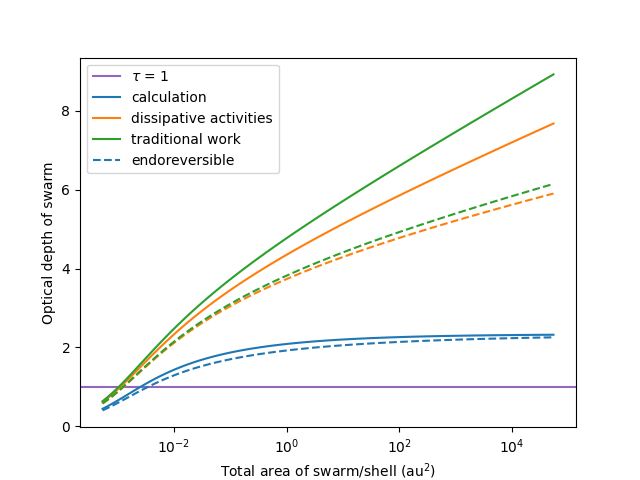}
    \caption{\textit{Left:} Optimum distance for a swarm of random orbiters with fixed total surface area for dissipative activities or traditional work, and for calculations, in the Landsberg limit and with $R_*=R_\sun$. The orbital distance of a shell of equal surface area is shown for reference. Values in the endoreversible limit are similar, and not shown. The optimum distance is largely insensitive to the kind activities done.  The solid shell radius is set by the surface area of a sphere. The random swarm increasingly would shadow itself at close distances but miss more flux at large distances; the optimum distance from the balance of these effects is typically about 40\% of the distance it would have as a shell for work and dissipative activities, and 70\% for computation. \textit{Right:} Optical depth of the swarms from the left panel, expressed as $\tau = A/(4\pi R^2)$, now including the endoreversible cases as dashed lines.  This ratio is 1 for the shell (shown in purple), although shells are assumed to have infinite optical depth.} 
    \label{fig:swarms}
\end{figure}

For computation, the optimum distance is somewhat different, being the maximum of equation \ref{eq:generalrate}, in particular the component
\begin{equation}
 \left[1-\exp{\left(-\frac{A}{4\pi R^2}\right)}\right]\left[\sqrt{\frac{R}{R_*}}-\left(\frac{R}{R_*}\right)^\frac{1-m}{2}\right]
\end{equation}
\noindent which is the computational rate in units of $\frac43 L/(kT_*\ln{(2)})$, and which can be greater than 1.

Both equations must be maximized numerically. The results are shown in Figure~\ref{fig:swarms}, compared with the radius of a complete shell of the same area, assuming $R_*=R_\sun$.

In both cases, the optimum distance for a swarm is quite similar to that of a shell for most distances---for dissipative activities and traditional work the distance is about 40\% of the shell distance, and for computation about 70\%.  The righthand side of that figure shows that for the swarm, significant shadowing takes place, with optimal optical depths of a few, depending on orbital distance.  This might argue that we should not expect Dyson spheres to be ``complete,'' but to provide a few magnitudes of gray extinction, regardless of their purpose.

\begin{figure}
    \centering
    \includegraphics[width=0.49\textwidth]{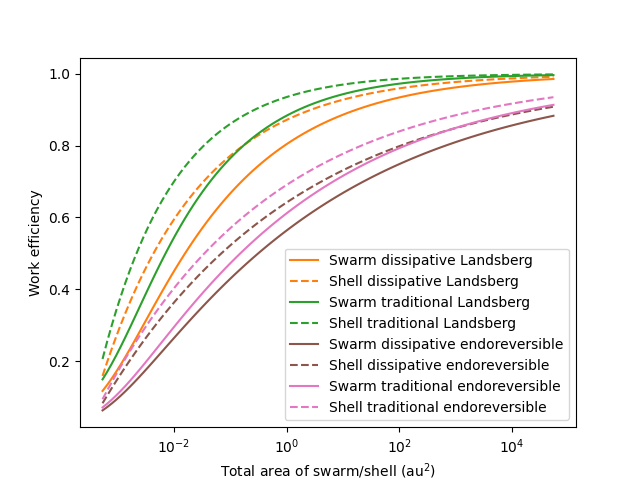}
    \includegraphics[width=0.49\textwidth]{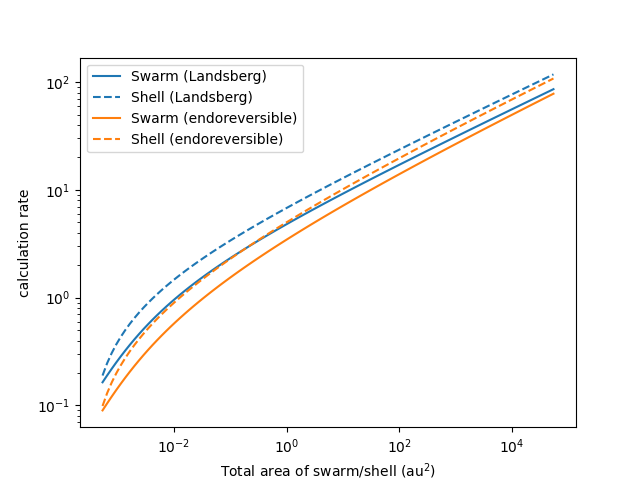}
    \caption{\textit{Left:} Efficiency of shells and swarms doing different kinds of activities in the Landsberg and endoreversible limits, assuming $R_* = R_\sun$. Note the plot is semi-logarithmic, and that all gains beyond 1 au$^2$ in surface area are no more than a factor of 2. \textit{Right:} Computational rate of shells and swarms in units of $\frac43 L/(kT_*\ln{(2)})$. For large areas, the total number of calculations scales very weakly as $r\propto A^\frac14$.}
    \label{fig:swarmseff}
 \end{figure}

The differences in efficiency of work and computation for the swarm versus the shell are not great, as shown in Figure~\ref{fig:swarmseff}.  In all cases, the efficiency does not vary by more than an order of magnitude for a huge range of total areas: there is very little gain per gram to be had once $\about 10\%$ of the starlight is being used, regardless of the assumptions we make.

\subsection{Multiple Shells}

So far we have optimized shells and swarms at a single distance, but there might be superior optima if the mass is spread out over some range of distances.  This is especially true if there is benefit to nested shells, with outer shells making use of the waste heat from inner shells, as in the idea of a Matrioshka Brain.

We first consider the case of two shells to see if inner shells offer any benefit over a single, outer shell.

\subsubsection{Computation}

First, we consider the case of computation in the Landsberg limit.  The inner shell, for which we will use the subscript $1$, does computation at a rate 
\begin{equation}
r_1 = \frac43 \frac{L}{kT_1 \ln{(2)}}\left[1-\frac{T_1}{T_*}\right]
\end{equation}
The second shell accepts the same amount of luminosity, and uses it to perform calculations at the rate
\begin{equation}
    r_2 = \frac43 \frac{L}{kT_2 \ln{(2)}}\left[1-\frac{T_2}{T_1}\right]
\end{equation}

\noindent so the total rate is $r_1+r_2$.  At the Landsberg limit, this is
\begin{equation}
r = r_1 + r_2 = \frac43 \frac{L}{kT_2 \ln{(2)}}\left[1-\frac{T_2}{T_*}\right]
\end{equation}

\noindent which is exactly the rate we would have had without the inner shell.  

This makes sense:  the total increase in entropy from the stellar surface to the outer shell is set only by their respective temperatures and radii---an outer shell that perfectly converts this entropy into computation is working at maximum efficiency. An inner shell might do some of these calculations and pass on the radiation to the outer shell to complete, but this merely changes the location of the increase in entropy from calculation.

Dyson spheres at the Landsberg and Landauer limits are thus neutral with respect to where the calculations take place.  If we imagine there are significant overhead costs to maintaining a sphere, such as for navigation, energy collection, and energy disposal, then the optimum configuration incurs these costs only in an outer shell, and does not bother with inner shells.

We can generalize this result to include practical limitations, as given by the endoreversible limit, by extending our analysis the the general $m<1$ case.  We compute the difference in computational rates $\Delta r$ of two shells versus one shell following the treatment above:

\begin{eqnarray}
        \Delta r &\propto& \frac{1}{T_1}\left[1-\left(\frac{T_1}{T_*}\right)^m\right] + 
                     \frac{1}{T_2}\left[1-\left(\frac{T_2}{T_1}\right)^m\right] - 
                     \frac{1}{T_2}\left[1-\left(\frac{T_2}{T_*}\right)^m\right] \\
                 &=& \frac{1}{T_2}\left[\frac{T_2}{T_1} - \frac{T_2}{T_1}\left(\frac{T_1}{T_*}\right)^m
                 - \left(\frac{T_2}{T_1}\right)^m  + \left(\frac{T_2}{T_*}\right)^m\right]\\
                 &=& \frac{1}{T_2}\left[\left(\frac{T_2}{T_1}\right)^m-\left(\frac{T_2}{T_*}\right)^m\right]\left[\left(\frac{T_2}{T_1}\right)^{1-m}-1 \right]
\end{eqnarray} 

The first term in brackets is never negative because $m\geq 0$ and $T_1<T_*$.  The second term is never positive because $T_2<T_1$ and $m\leq1$.  When $m=1$ in the Landsberg limit, we recover the result that $\Delta r =0$, that is that there is no benefit or harm to inner spheres beyond additional overhead per sphere.  For other values of $m$, inner shells only decrease overall computational efficiency.

Thus, there is no reason at the Landauer limit to build inner shells, regardless of the efficiency model we assume for the engines providing energy to the computers. Matrioshka brains are not computationally ideal.

\subsubsection{Dissipative Activities}

For dissipative activities internal to the system, the two shells extract work
    
\begin{equation}
\Winternalone = L \left[1- \left(\frac{T_1}{T_*}\right)^m\right]
\end{equation}

\begin{equation}
\Winternaltwo = L \left[1- \left(\frac{T_2}{T_1}\right)^m\right]
\end{equation}

\begin{equation}
\Winternal = \Winternalone+\Winternaltwo = L \left[2 - \left(\frac{T_1}{T_*}\right)^m - \left(\frac{T_2}{T_1}\right)^m\right]
\end{equation}

\noindent which indicates an efficiency greater than one.  Here, we see there can be a benefit to additional shells: each provides additional internal work that can be done, at efficiencies near unity.  There is a limit, however.  If we imagine an infinite number of shells between two temperatures $T_h$ and $T_c$, the total \Winternal\ is
\begin{eqnarray}
    \Winternal &=& L\left[1-\left(\frac{T_h}{T_*}\right)^m + \int_{T_c}^{T_h} \left(1-\left(\frac{T-dT}{T}\right)^m\right)\right]\\
    &=& L\left[1-\left(\frac{T_h}{T_*}\right)^m + m\int_{T_c}^{T_h} \frac{dT}{T}\right]\\
    &=& L\left[1-\left(\frac{T_h}{T_*}\right)^m + m\ln{\left(\frac{T_h}{T_c}\right)}\right]
\end{eqnarray}

\noindent If we guess at some extreme values to bound the utility of so may shells, $T_* = 6000$ K, $T_h = 1000$ K, $T_c = 10$ K, we have at the Landsberg limit ($m=1$)
\begin{equation}
    \Winternal \approx L (0.83 + 4.6) \approx 5.4 L
\end{equation}

Building five spheres with temperatures $T_i = 1,000$K, 300 K, 100 K, 30 K, and 10 K yields $\Winternal = 3.56 L$, which is over half of the benefit of an infinite number of spheres.  We thus see some benefit to a set of up to several nested spheres at very different temperatures, with diminishing returns, which decreases somewhat as $m$ does.

The optimal arrangement of these shells as a function of available mass is a complex problem that probably does not warrant too detailed a solution since we do not know what practical effects we are missing.

The lesson is that for such activities at or near the Landsberg limit we might expect a swarm of material at many distances, with the outer layers exploiting the waste heat of the inner layers.  This might also be true for computation if it occurs well above the Landauer limit in a manner where $r$ is independent of temperature but proportional to \Winternal, and so might justify a Matrioshka brain.

\subsubsection{Traditional Work at the Landsberg limit}

We look now to work that leaves the system.  Computing the total work for the first shell we have:
\begin{eqnarray}
    \dot{W}_1 &=& L \left[1- \left(\frac{T_1}{T_*}\right)^m\right]\\
\dot{W}_2 &=& L \left(\frac{T_1}{T_*}\right)^m \left[1- \left(\frac{T_2}{T_1}\right)^m\right]
\end{eqnarray}

\noindent where the second shell can only access the remaining energy that has not left the system from the first shell.  Summing these two, we have
\begin{equation}
\dot{W} = \dot{W}_1 + \dot{W}_2 = L \left[1- \left(\frac{T_2}{T_*}\right)^m\right]
\end{equation}

\noindent which is again exactly what we would have had if we had not bothered with the first shell, regardless of the efficiency model we adopt.  Dyson spheres are thus also neutral to where the work is performed, but simplicity might dictate all work is optimally performed in a single, outer shell.

\subsection{Some Other Practical Effects}

There are many practical limitations that will change our results, mostly having the effect of reducing the efficiencies of Dyson spheres below the limits here.

Two small effects that may actually make our limits here \textit{pessimistic} are the actual spectrum of the star and the limitations of optical circulators.  

Because real stellar spectra are not blackbodies, their outgoing flux has \textit{less} entropy per unit energy than we have assumed when we interpret their effective temperature $\teff$ as $T_*$.  This ultimately means that one can extract more work or do more computations than we have assumed, depending on the degree of departure from a blackbody.  

We have invoked optical circulators to ignore energy that falls back onto the star from the Dyson sphere. Including this must be done with care, but for now we will ignore changes to the star's structure and focus on the first-order heating effect on the surface. When such energy lands on a star with temperature $\teff \equiv (L/(4\pi R_*^2\sigma))^\frac14$, the surface heats to a new, higher temperature $T_*$, and returns this energy to the shell, and this must be accounted for explicitly. This new temperature, by energy balance, obeys
\begin{equation}
    T_*=\left(\teff^4 + T^4\right)^\frac14 = \teff\left[1+\left(\frac{T}{\teff}\right)^4\right]^\frac14
\end{equation}

\noindent where $T$ is the temperature of the shell.

This has no effect on the outer temperature of the shell for computation and dissipative activities, which is set entirely by $L$ and $R$.

The radiation the shell receives back from the star, however, is at a higher temperature and so has less entropy per erg than it had going in.\footnote{This does not violate the second law of thermodynamics since energy gains entropy as it moves from the stellar core to the surface, and this process can become slightly less effective to compensate for the extra entropy arriving from the outside.} This slightly lowers the entropy flux through the shell, by a geometric factor of order $1+(T/\teff)^4$. Since the temperature of a shell doing traditional work is set by entropy balance, this slightly lowers its temperature.

For calculations, which work directly with the entropy of the radiation, the result of all of this is an additional amount of computation that can be done, introducing a correction of order $r(T/\teff)^3$ (since $r$ scales as $L/T$ times the geometric factor).  This is thus, at best, perhaps a few percent for a very warm sphere.  The amount of dissipative activities and traditional work go as $L(1-\frac34 \Sdotin T)$, so decrease by a small amount because \Sdotin\ is higher.

For swarms with finite optical depth, not using circulators has an additional benefit that it allows them to cool via inward radiation into deep space (i.e.\ via paths 2a and 3a in Figure~\ref{fig:geometry}). This will not affect their ability to achieve the Landsberg limit in terms of $T$, but it will lower $T$ by increasing their available radiating surface to nearly the full sphere, i.e.\ by almost a factor of 4.  The details will depend on the swarm geometry in complex ways, but the extra cooling will very roughly be a factor of $\about(1 + 3e^{-\tau})$. For complete spheres, this will increase the swarm performance levels we have calculated, closer to the shell performances, and move the optimal distances outwards. For dense swarms the improvement is small but for our Landsberg computation case where optimal optical depths are of order 2, it could plausibly raise $r$  by over 10\%. In the $\tau{\ll}1$ limit where we consider only isolated swarm elements, optimal values of $r$ will be higher by up to a factor of 2 or so. 

These (and doubtless many more effects) are all likely much smaller than the uncertainties introduced by our assumptions and approximations elsewhere, but could be considered in any detailed analysis of a specific design of Dyson sphere. 

\section{Discussion and Conclusions}

The details of optimal ways to arrange mass in a Dyson sphere depend on assumptions about the nature of the engines.  Some overall themes have emerged from our analysis, however:
\begin{itemize}
    \item Unless one has enough mass to capture most of a star's luminosity, the optimal placement of mass is as close to the star as possible, favoring higher intercepted flux over efficiency.  This argues that unless a star is suffering significant optical extinction, we should expect the waste heat of industry to be in the mid- or near-infrared, or even in the optical (a possibility explored by \citet{Osmanov18} and \citet{Lacki16}.)
    \item Even for complete spheres, the return on investment for additional amounts of mass beyond a small sphere is extremely small, also arguing for relatively high sphere temperatures.
    \item In principle, optical circulators could be used to avoid complications of feedback among elements of a Dyson swarm. Activities might also be done directly with photons, avoiding the need to run heat engines between intermediate absorbers heated and cooled by radiation. 
    \item In that ultimate limit, the appropriate optimal (Landsberg) efficiency for Dyson sphere components is the Carnot efficiency.  For computation there is an additional factor of $1/T$ to consider from the Landauer limit.
    \item Unless they radiate their energy out of the system, for instance as low-entropy radio waves, Dyson spheres as a whole do no ``work'' in a thermodynamic sense, since they must radiate away all energy they consume as waste heat. This raises their temperatures and lowers their efficiencies compared to engines that do traditional work.
    \item The optimal orbital distances as a function of mass have some dependence on the details of the kind of activities performed and the nature of the engines, but these details only matter to  a factor of 2 or so.
    \item If Dyson spheres are composed of swarms capturing most of the star's light, they likely have optical depths of a few.
    \item For work that leaves the sphere and computation at the Landauer limit, there is no value to nested spheres, each capturing the inner sphere's waste heat. Matrioshka Brains are thus not optimal configurations for performing calculations. 
    \item For dissipative activities, such as maintaining biological activity or computation below the Landauer limit, there is some benefit to widely spaced nested spheres feeding on each others' waste heat.  In these cases, we might expect material across a wide range of orbital distances, and the total optical depth to the star could be quite high.
\end{itemize}

We must, of course take all of this with a grain of salt. Real technological development around a star will be subject to many constraints and practical considerations that we probably cannot guess.  While we have outlined the ultimate physical limits of Dyson spheres, consistent with Dyson's philosophy and subject only to weak assumptions that there is a cost to acquiring mass, if real Dyson spheres exist they might be quite different than we have imagined here.

Nonetheless, these conclusions can guide speculation into the nature of what sorts of Dyson spheres might exist, help interpret upper limits set by search programs, and potentially guide future searches.

\begin{acknowledgements}
I thank Viorel Badescu for feedback on and suggestions for this manuscript. I thank Caleb Scharf and Brian Lacki for helpful conversations. The Center for Exoplanets and Habitable Worlds and the Penn State Extraterrestrial Intelligence Center are supported by Penn State and its Eberly College of Science. This research has made use of NASA's Astrophysics Data System Bibliographic Services.
\end{acknowledgements}

\bibliographystyle{aasjournal}

\end{document}